\documentclass[preprint,11pt]{JHEP3}

\JHEPspecialurl{http://jhep.sissa.it/JOURNAL/JHEP3.tar.gz}

\usepackage{epsfig,multicol,amsmath}

\usepackage{bm}  
\usepackage{amsmath}     
\usepackage{epsfig,epsf}
\usepackage{cite}
\def\beq{\begin{equation}}
\def\eeq{\end{equation}}
\def\bea{\begin{eqnarray}}
\def\eea{\end{eqnarray}}

\def\eq#1{{Eq.~(\ref{#1})}}
\def\fig#1{{Fig.~\ref{#1}}}
\newcommand{\bas}{\bar{\alpha}_S}

\newcommand{\Lb}{\left(}
\newcommand{\Rb}{\right)}
\newcommand{\h}{\frac{1}{2}}
\parskip=\itemsep               

\newcommand{\nn}{\nonumber}

\newcommand{\ga}{\gamma}

\newcommand{\om}{\omega}

\def\pom{{I\!\!P}}

\title{Gluon saturation: survival probability for leading neutrons in DIS}
\author{\Large 
Eugene\, Levin${}^{a, b}$ \thanks{Email: leving@post.tau.ac.il, eugeny.levin@usm.cl.}\,\,\,and\,\,\, Sebastian Tapia${}^b$ \thanks{Email: trockut@gmail.com}
\\
${}^a$ \, Department of Particle Physics, School of Physics and Astronomy,
Tel Aviv University, Tel Aviv, 69978, Israel\\
${}^b$\, Departamento de F\'\i sica,
Centro de Estudios Subat$\acute{o}$micos,
Universidad T$\acute{e}$cnica Federico Santa Mar\'\i a,\\ and
Centro Cientifico-Tecnol$\acute{o}$gico de Valpara\'\i so,
Casilla 110-V,  Valparaiso, Chile\\
}

\abstract
{In this paper  we discuss the example of one rapidity gap process: the  inclusive cross sections of the leading neutrons in deep inelastic scattering  with protons (DIS). The equations  for this process are proposed and solved, giving the example of theoretical calculation of the survival probability for one rapidity gap processes.  It turns out that the value of the survival probability is small and it decreases with energy. }

\keywords{Color Glass Condensate, gluon saturation, BFKL Pomeron calculus,  non-linear evolution, geometric scaling behavior
}
\dedicated{PACS: 12.38-t, 12.38.Cy,1 2.38.Lg, 13.60.Hd, 24.85.+p, 25.30.Hm}
\preprint{TAUP 2929/11 \\
{\tt }\\
\today}

\begin{document}
\section{Introduction.}
The inclusive cross sections of the leading neutrons in deep inelastic scattering with protons (DIS) attracts  interest since  this process allows us to extract the $F_2$ deep inelastic structure function for pions ($F^\pi_2$) (see \fig{genpic} and Refs.\cite{BA} ). This process has one rapidity gap since no particles are produced in the region of rapidity between the neutron and the bunch of secondary particles with mass ($M_X$ in \fig{genpic}). For long time it has been known that  the cross section of such processes has to be multiplied by factor  $S^2$ which is called survival probability\cite{BJ,DOK,GLM1}. This factor stems from the possible interaction of the constituents of the projectile with the target that should be forbidden to preserve the gap.
In other words, the constituent of projectile could interact with the target in initial or final state suppressing the cross section of such process (see \fig{gensc} that illustrates such interactions). 

\FIGURE[h]{
\begin{tabular}{c }
\epsfig{file=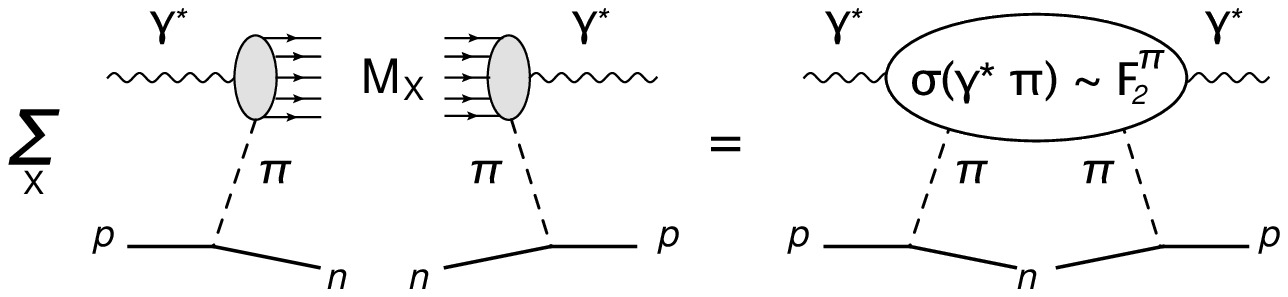,width=100mm}\\
\end{tabular}
\caption{The generic picture for leading neutron production in DIS: the Born approximation.)}
\label{genpic}
}
The two decades experience in calculation of the survival probability ( see Refs.\cite{KMRS,GLMSP,BGR,BO1,NNN,DA,MORI}) has shown that the predictions for survival probability could differ in the orders of magnitude (from  $S^2$  = 0.9 to 0.01) depending on the model for strong interaction at high energy.

In this paper we develop the theoretical approach for survival probability in DIS based on high density QCD
\cite{GLR,MUQI,MV,JIMWLK,B,K}.  For DIS at high energy  QCD predicts that the dynamics of quarks and gluons (partons) can be described in terms of  parton saturation\footnote{Parton saturation means that at high energy the density of partons (actually mostly gluons) reaches the maximum value. It is instructive to notice that  in CGC  the QCD coupling is small ($\bas \ll 1$). }\cite{GLR,MUQI,MV}. The new state of the matter: Color Gluon Condensate (CGC)\cite{MV,JIMWLK} will be produced in which
the amplitudes are dominated not by quantum fluctuations, but by the configurations of classical field containing large, $\sim 1/\bas$ numbers of gluons. This state is characterized by large density but small QCD coupling $\bas$. This feature results in the solid theoretical approach based on non-linear equations \cite{GLR,MUQI,JIMWLK,B,K}. It worthwhile mentioning that this theoretical approach has reached the most developed stage for DIS with which we are dealing in this paper.

\FIGURE[h]{\begin{minipage}{8cm}{
\centerline{\epsfig{file=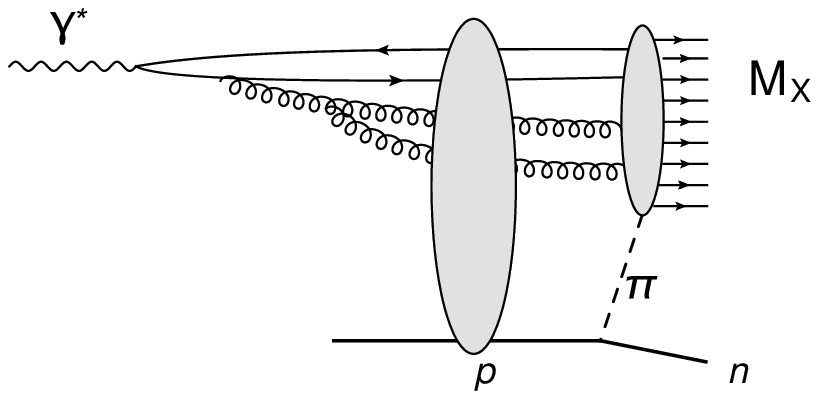,width=75mm}}}
\end{minipage}
\begin{minipage}{6cm}{
\caption{Leading neutron production in DIS:  general source of shadowing corrections }
\label{gensc} }
\end{minipage}
}
~
The main result of the paper is the equation for the inclusive production of leading neutron. The basic theoretical ideas how to approach the processes that have the rapidity gap, have been formulated in Ref.\cite{KL}. We explore these ideas to obtain the equation. We found the solution to the new equation and show that the interaction in initial and final sate will lead to the cross section that falls down at high energy. In other words, the survival probability turns out to be small at high energy. 

The paper is organized as follows. In the next section we  will derive the equation and discuss the high energy asymptotic behaviour of the solution. Section 3 is devoted to the initial condition to the equation, while in section 4 we obtain the numerical solution to the equation. In conclusion we summarize the main results.

\section{The equation}

\subsection{Derivation of the equation in the dipole approach}


In inclusive leading neutron production (ILNP) the produced partons could interact between themselves as well as with the target in the final state, unlike in the case of total cross section for which we have the non-linear Balitsky-Kovchegov  equation \cite{B,K} . In principle, the interaction in the final state means that the cross section of the process does not depend only on the structure of the wave function of the colliding particle as it is  for the total cross section. However, in Ref.\cite{KL}(see also Ref. \cite{KOVD})   it is shown on the example of diffraction production, that in the processes which are inclusive, the different type of interactions in the final state cancel each other. In this section we argue that ILNP is also belongs to the class of such inclusive processes as diffraction production.

In ILNP we can see three typical moments of time\footnote{We will consider all processes in the light-cone frame and will use the light-cone perturbative theory (see Ref.\cite{BRODLE})} :  at $x^- = - \infty$ we have fast virtual photon
(or, better to say, a system of partons in the coherent state of the virtual photon) and the target, at $x^- \,=\,0$ these
partons  interact with the target and the produce   partons that propagate to $x^-= + \infty$ where the detectors are placed which measure this system of partons. Actually in our case we do not measure these partons which means that we sum over all possible final states with only one restriction that their total mass is equal to $M_X$(see \fig{genpic}). The same time structure we have in complex conjugated amplitude in which we denote times as $\bar{x}^-$ (see \fig{timestr}).


\FIGURE[h]{
\begin{tabular}{c}
\epsfig{file=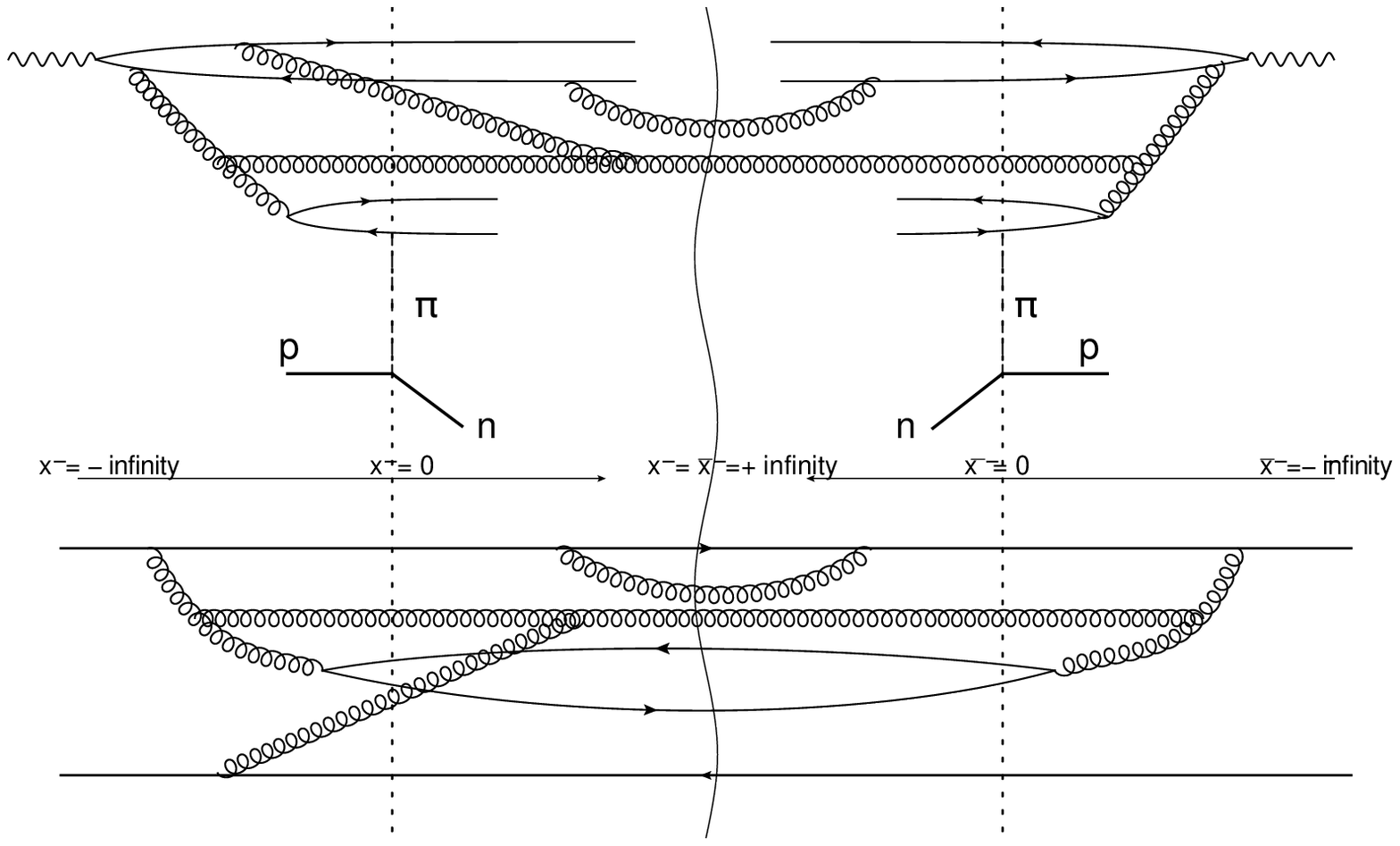,width=100mm,height=50mm}\\
\end{tabular}
\caption{The time structure of the inclusive production of leading neutrons.The low panel shows  our notation for the process shown in the upper panel. The straight solid lines denote the initial dipole in the low panel. }
\label{timestr}
}


\fig{timestr} shows the particular contribution to the cross section of interest. From $x^- = - \infty$ to $x^- =0$ the wave function of the fast dipole consists of system of two quark-antiquark pairs and two gluons, at   $x^- = 0$ this system interacts with the pion which destroys its coherence. As a result the components of the system (two gluons and two quark-antiquark pairs ) are produced.  The components  interact in  the time interval $0   <   x^- < + \infty $.  The gluons can be emitted and absorbed in the final state. Because of this we cannot use optical theorem that reduces the cross section to the imaginary part of the elastic amplitude.  The key observation is that absorption and emission in the final state ( for times $x^- >0$ and $\bar{x}^- > 0$) cancel each other. This cancellation was first proven in Ref.\cite{CHMU}.  It is found  that  we have two types of cancellations 
shown in \fig{2cancel}. The first type is cancellation of the gluon emitted  in the final sate and   which can be caught  by the detector (see the  first diagram in \fig{2cancel}-A) with gluons emitted and absorbed in the final state in the amplitude and the conjugated amplitude (see the second and third diagrams in \fig{2cancel}-A). The second type of cancellation is shown in \fig{2cancel}-B. The produced gluon from the initial wave function, that exists  from $x^- = -\infty$ to $x^- =+ \infty$,  cancels by the gluon from the wave function that has been absorbed in the final state and, therefore cannot be measured by the detector. Notice that the gluon in  the first diagram of \fig{2cancel}-B has been emitted in the final state in the conjugated amplitude. In \fig{2cancel} we denote the gluon by  quark and antiquark lines since we consider our process at large $N_c$ where $N_c$ is the number of colours.

\FIGURE[h]{
\begin{tabular}{c}
\epsfig{file=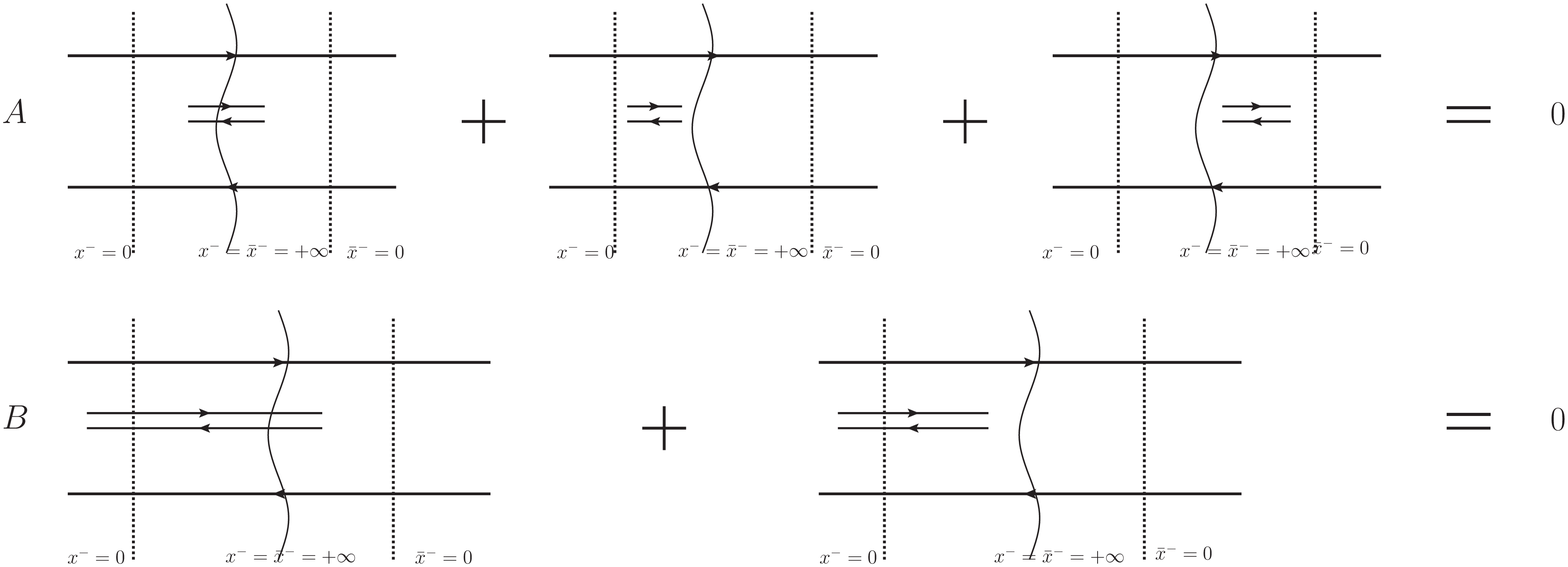,width=100mm,height=50mm}\\
\end{tabular}
\caption{Two types of cancellations in the final state}
\label{2cancel}
}


 Armed with these cancellations we can show that in our process the emission and absorption in the final state does not contribute to the cross section. To illustrate this point we consider the example of \fig{timestr}.  One can see from \fig{cantime} that emission of the  upper gluon in \fig{cantime}-1   is canceled by diagram of \fig{cantime}-2 which describes the emission and absorption of the upper gluon in the final state (actually for cancellation we need to add the diagram of \fig{cantime}-2 type for the conjugated amplitude). The absorption of the low gluon in \fig{cantime}-1 is canceled by the diagram of \fig{cantime}-3.  This diagram describes the emission of the gluon   from the initial state in the amplitude but the emission of gluon from the final state in the complex conjugated amplitude.  The cancellation of emission and absorption in the final state means that we can used the optical theorem for interaction of the dipole with virtual pion (see \fig{genpic}).  It is worthwhile mentioning that this cancellation works in any  complex diagrams which differs only by the gluon shown in \fig{2cancel} \cite{CHMU}.
 
 The formal proof of the discussed cancellation can be done using the method of mathematical induction. Let us assume 
 that for emission of $n$-gluons we have the cancellation and they can be characterized  by the initial wave function of colorless dipole.  The emission of $n + 1$ gluon by one dipole can be described by the sum of the diagrams of \fig{1gl}. One can see that diagrams of the set A in \fig{1gl} cancels as well as the diagrams of set B , due to the cancellation give by \fig{2cancel}.
 The only diagram that remains and gives the contribution is the diagram of \fig{1gl}-C.
 Since the initial condition for $N^\pi$ contains only one dipole we conclude that we can describe this processes by the initial wave function of dipoles.

Now we can derive the equation based on the same ideas as have been used in the derivation of Balitsky-Kovchegov (BK) equation, since the ILNP stems from the structure on the initial wave functions of colliding particles. Let us denote by $N^\pi\Lb x, b,Y,Y_g\Rb $ the imaginary part of the 
amplitude shown in \fig{genpic}. $Y_g$ denotes the rapidity gap between the neutron and the slowest  hadron  among the produced particles (see
\fig{genpic}).


\FIGURE[h]{
\begin{tabular}{c}
\epsfig{file=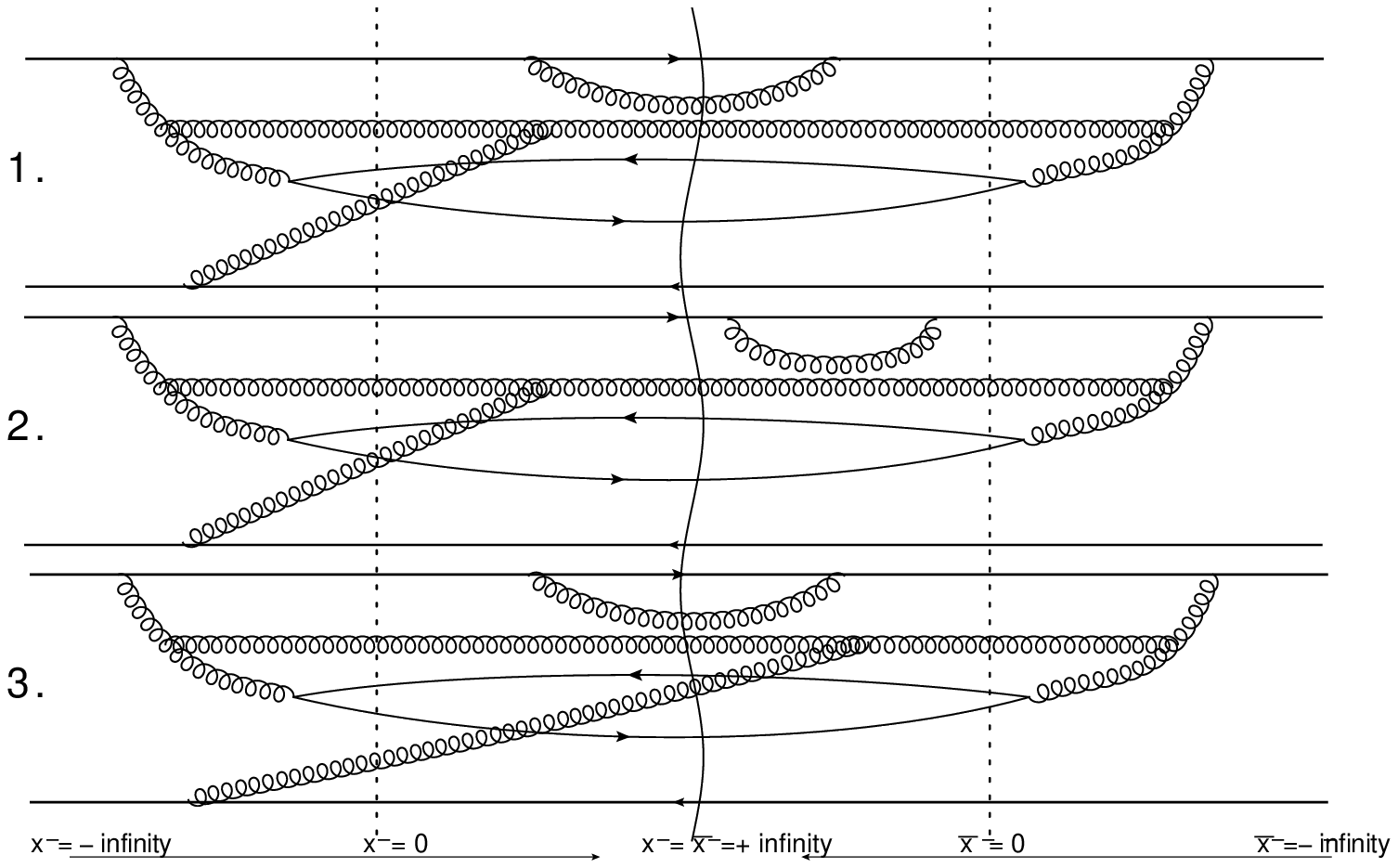,width=100mm}\\
\end{tabular}
\caption{The example of cancellation of emission and absorption in the final state for the contribution of the diagram of \fig{timestr} to the cross section.}
\label{cantime}
}


A long before the interaction the incoming dipole decays into two dipole each of them could scatter separately and produced the leading neutron. These two dipoles can interact simultaneously producing leading neutrons. However, we have another process which does not produced any particles that fill the rapidity gap. One dipole scatters elastically while the second produces the leading neutron.  \fig{eq} shows these  three ways of interaction. All these terms we can see in \fig{eq}. The first term shows the separate interaction of each dipole while the second term describe the 
simultaneous interaction of two dipoles. The third and the forth terms describe the ILNP by one dipole while the second dipole undergoes the elastic scattering. The analytical form of this equation looks as follows:

\FIGURE[h]{
\begin{tabular}{c}
\epsfig{file=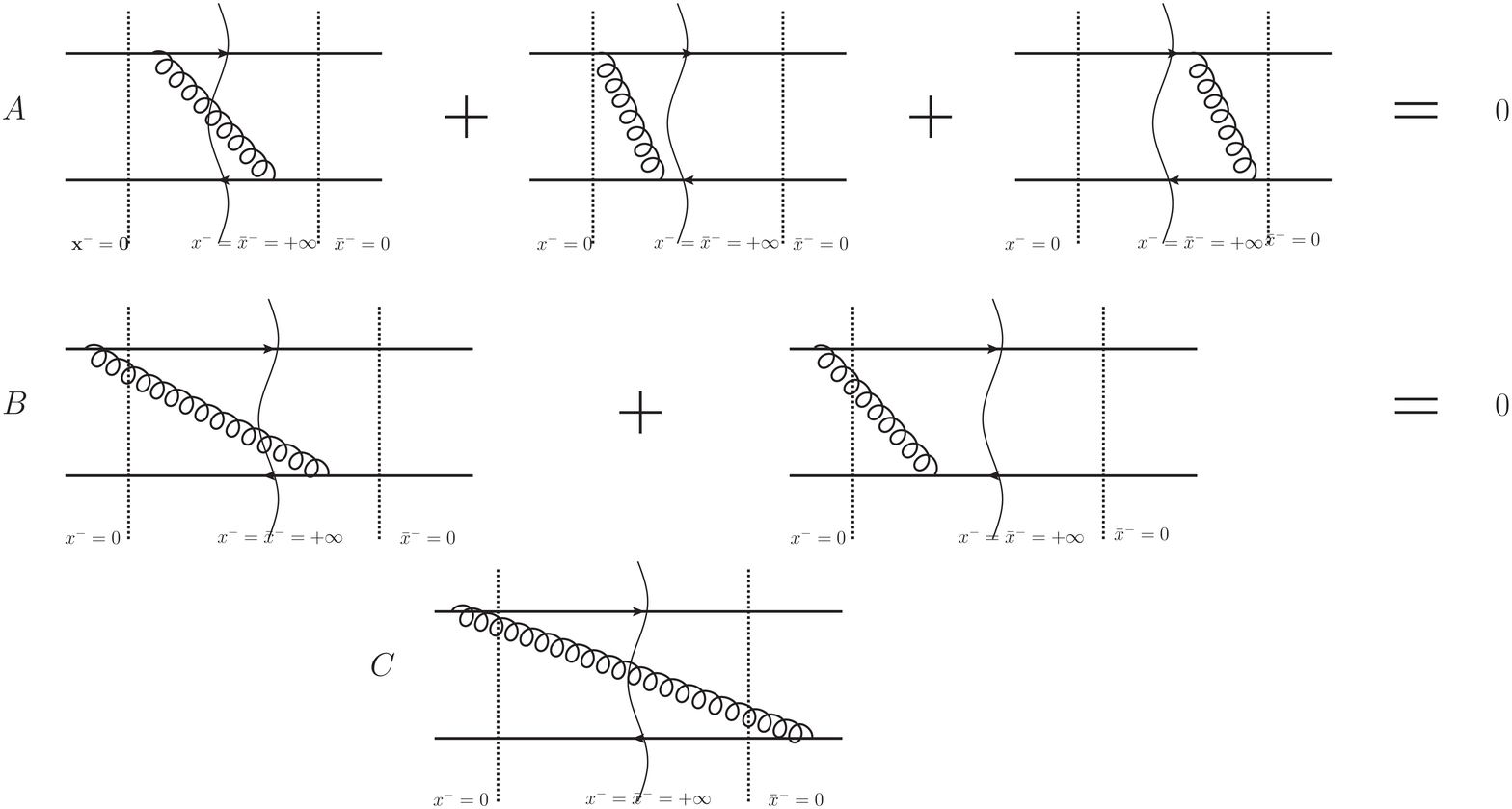,width=120mm}\\
\end{tabular}
\caption{The emission of one gluon by the dipole. The diagrams with the gluon emitted and absorbed by the same quark (antiquark) are not shown but they have the same  pattern.}
\label{1gl}
}

\FIGURE[h]{
\begin{tabular}{c}
\epsfig{file=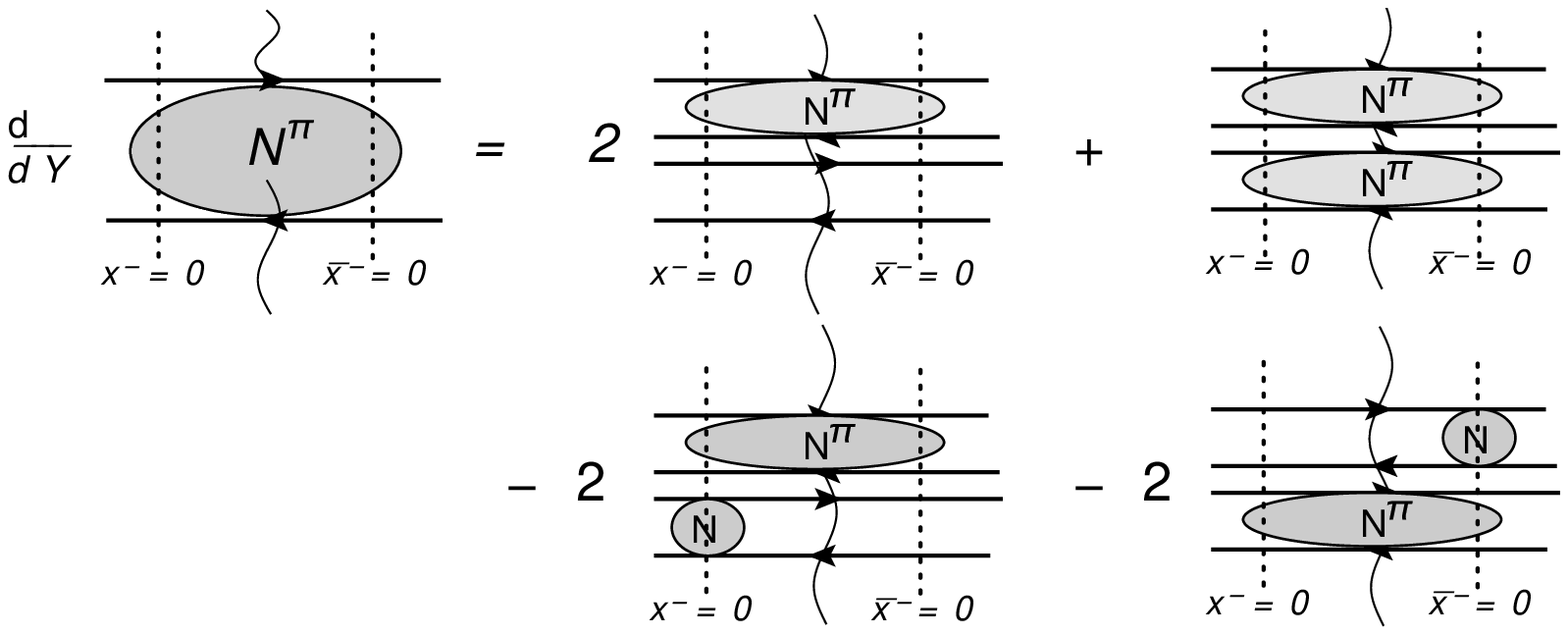,width=100mm}\\
\end{tabular}
\caption{The  graphic form of the equation. The vertical dotted lines denote the time of interaction in the amplitude or in complex conjugated amplitude.
If such line crosses the blob it means that the interaction occurs. The wave lines denote the infinite time and if such line goes through the blob it means that the blob describe the production of particles.}
\label{eq}
}

\bea \label{EQ}
&&\frac{\partial N^\pi\Lb x_{10}, b;Y,Y_g\Rb}{\partial Y}\,\,=\\
&=&\,\,\frac{\bas}{2 \pi}\,\int d^2 x_2 \frac{x^2_{10}}{x^2_{02}\,x^2_{12}}\,\left\{N^\pi\Lb x_{02}, \vec{b} - \h\vec{x}_{12};Y,Y_g\Rb\,+\,N^\pi\Lb x_{12}, \vec{b} - \h\vec{x}_{02};Y,Y_g\Rb\right.\nn\\
&-& \,\,\left.N^\pi\Lb x_{12}, \vec{b} ;Y,Y_g\Rb\,\,+\,\, N^\pi\Lb x_{02}, \vec{b} - \h\vec{x}_{12};Y,Y_g\Rb\,N^\pi\Lb x_{12}, \vec{b} - \h\vec{x}_{02};Y,Y_g\Rb\right.\nn\\
 &-&\,\,\left.2  N^\pi\Lb x_{02}, \vec{b} - \h\vec{x}_{12};Y,Y_g\Rb\,N\Lb x_{12}, \vec{b} - \h\vec{x}_{02};Y \Rb\,\,-\,\,2  N\Lb x_{02}, \vec{b} - \h\vec{x}_{12};Y \Rb\,N^\pi\Lb x_{12}, \vec{b} - \h\vec{x}_{02};Y,Y_g\Rb \right\}\nn
  \eea
where $N\Lb x_{10},b,Y\Rb$ is the solution of  Balitsky-Kovchegov equation.

\subsection{ BFKL Pomeron calculus and inclusive production of leading neutron}

As it is well known the non-linear BK equation for the elastic amplitude corresponds to summation of  `fan' diagram in the framework of the BFKL Pomeron calculus \cite{GLR,BRA}. The equation for the ILNP stems from the AGK cutting rules\cite{AGK}. These rules has been proven in QCD for all processes in which we do not have emission of the gluon from the triple BFKL Pomeron vertex (see Refs.\cite{LEPR,JAKO,KOTU,GEVE,BARY,TRE}). Our process belongs to this class and we can safely apply the AGK cutting rules.  The contribution to the imaginary part of the elastic scattering amplitude can be viewed as sum to three different processes which are shown in \fig{agkproc}. If the rapidity $Y'$ is the position of the triple Pomeron vertex the simplest `fan' diagram generates three different processes: in the first one (double cut in \fig{agkproc})  the state with double density of particle(gluons) is produced while in the second process (single cut in \fig{agkproc}) the density of the particle is the same as in the single Pomeron exchange; the third process is the diffraction production in which we do not produce  a particle in this rapidity gap. The AGK cutting rules establish the relations between these  processes and they are shown in \fig{agk}-A.


\FIGURE[h]{
\begin{tabular}{c c}
\epsfig{file=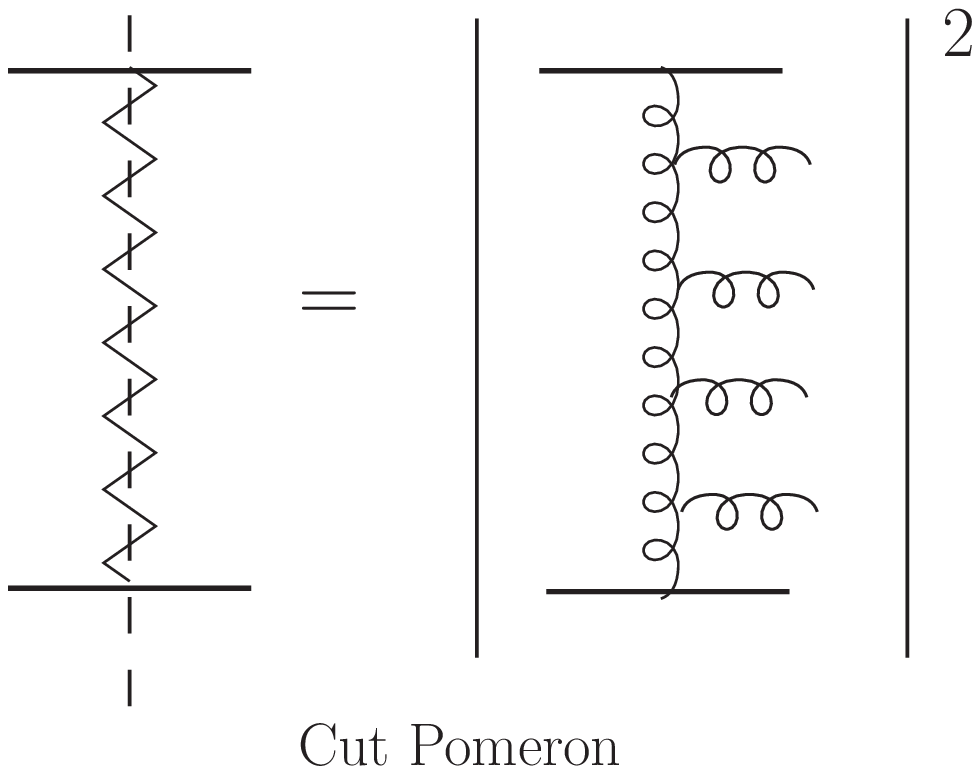,width=40mm}& \epsfig{file=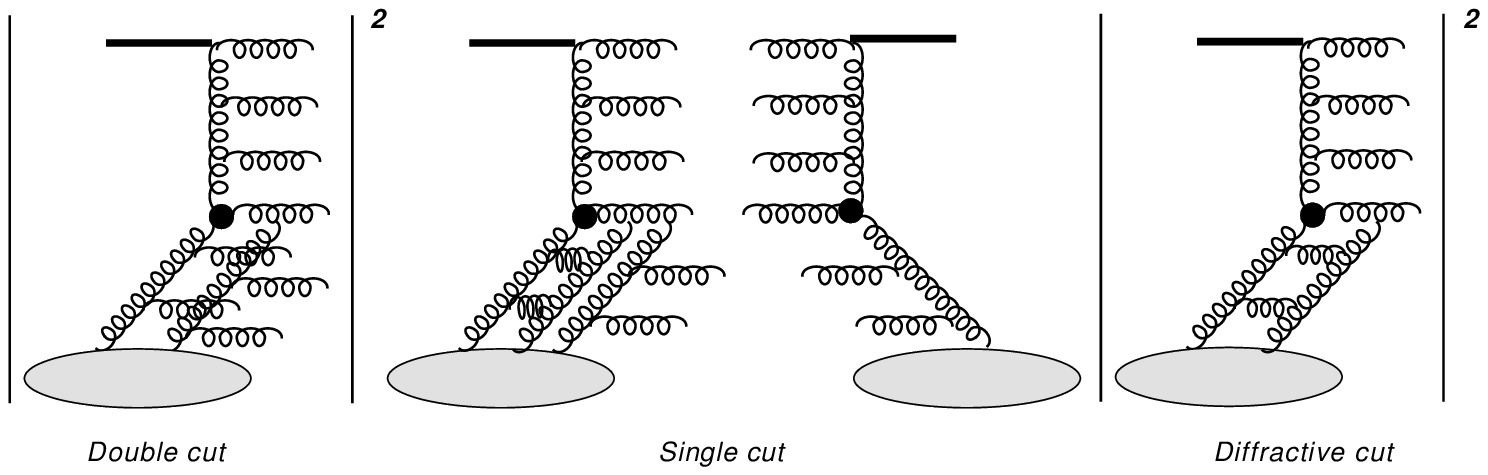,width=90mm}\\
\fig{agkproc}-a & \fig{agkproc}-b\\
\end{tabular}
\caption{The processes that corresponds to triple BFKL Pomeron contribution to the total cross section..}
\label{agkproc}
}

\FIGURE[h,b]{
\begin{tabular}{c }
\epsfig{file=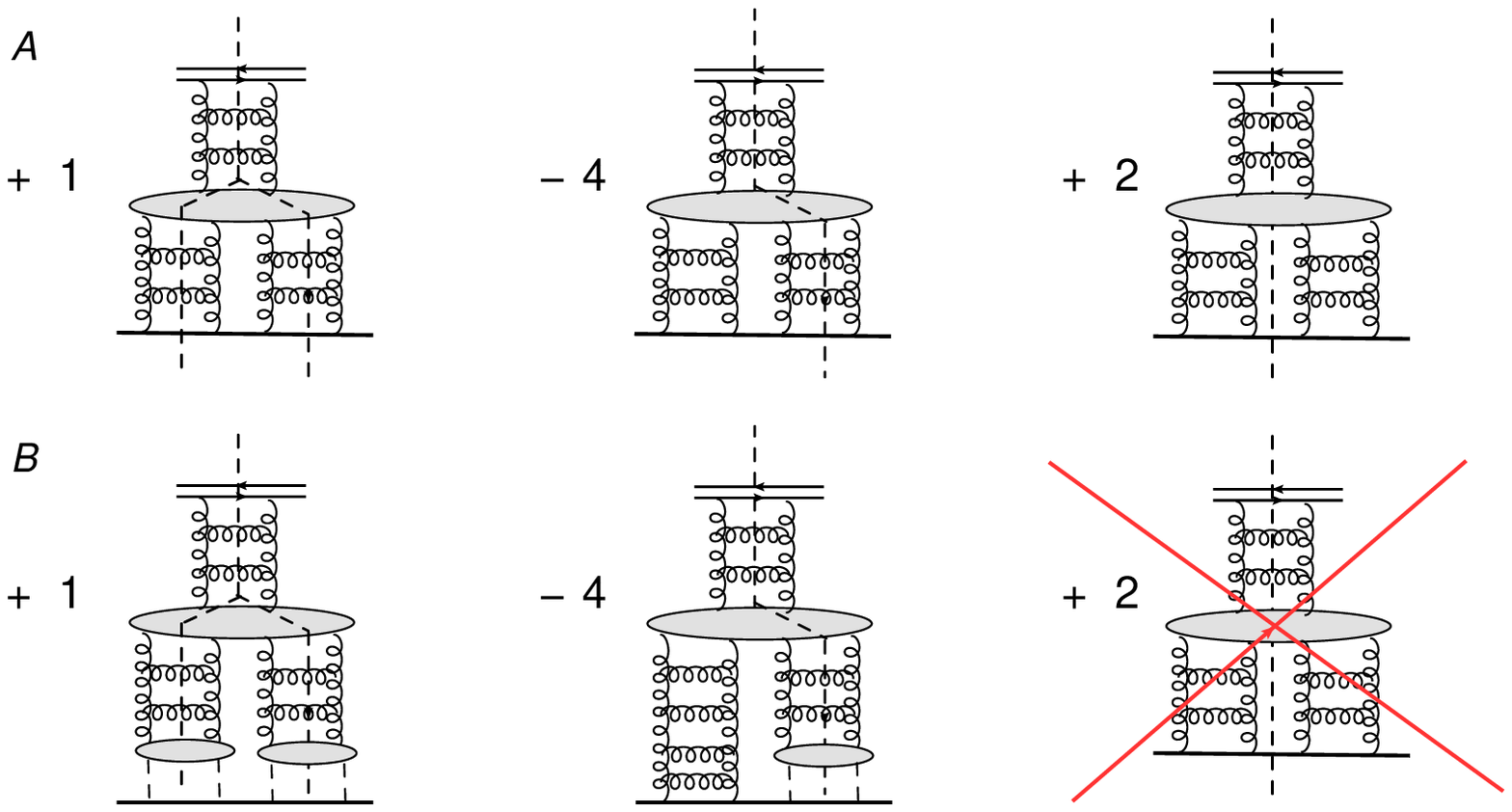,width=90mm}\\
\end{tabular}
\caption{The relation between the AGK cutting rules and contribution to the inclusive production of leading neutrons. The helix lines denote  gluons, pions are shown by the dashed lines. It should be noticed that in \fig{agk}-B between two exchanges of pions the thick line denotes the neutron.}
\label{agk}
}

The difference between coefficients that are shown in \fig{agk}-A and the standard ones is related to the fact that we use  for the contribution of the cut Pomeron the contribution to the total cross section (see \fig{agkproc}-a). instead of the contribution to the imaginary part of the amplitude. As we know from the unitarity constraint:   $ 2\, \mbox{Im}\,A \,=\, \sigma_{tot}$.

Having \fig{agk}-A in mind we can see that the leading neutrons can be produced from the first and second contributions while the third one corresponds to the diffraction production and it contributes only to the spectrum of leading protons (see \fig{agk}-B). Replacing Pomerons by the the sum of `fan' diagrams we obtain the equation given by \eq{EQ}. The direct relation between  dipole language in QCD and the BFKL Pomeron calculus for the inclusive processes have been noticed in Ref.\cite{KL}. The powerful theorem on cancellation of the interaction in the final state in the Pomeroin language are hidden in the assumption that only Pomerons and their interaction contribute to the inclusive processes.  Therefore, the dipole consideration can be considered as the theoretical argument for the Pomeron calculus.

\subsection{Solution at ultra high energy}
We need to find an initial condition to \eq{EQ} before searching for the solution. However, the experience with BK equation shows that the asymptotical behaviour at  high energies does not depend on the initial solution \cite{LT}.  Using the approach of Ref. \cite{LT} we can find the solution noticing that the solution to BK equation approaches unity at ultra high energies. Substituting $N =1$ in \eq{EQ} we obtain the following equation for $N^\pi$

~

~

\bea \label{EQHE}
&&\frac{\partial N^\pi\Lb x_{10}, b;Y,Y_g\Rb}{\partial Y}\,\,=\\
&=&\,\,\frac{\bas}{2 \pi}\,\int d^2 x_2 \frac{x^2_{10}}{x^2_{02}\,x^2_{12}}\,\left\{-\,N^\pi\Lb x_{02}, \vec{b} - \h\vec{x}_{12};Y,Y_g\Rb\,-\,N^\pi\Lb x_{12}, \vec{b} - \h\vec{x}_{02};Y,Y_g\Rb\right.\nn\\
&-& \,\,\left.N^\pi\Lb x_{12}, \vec{b} ;Y,Y_g\Rb\,\,+\,\, N^\pi\Lb x_{02}, \vec{b} - \h\vec{x}_{12};Y,Y_g\Rb\,N^\pi\Lb x_{12}, \vec{b} - \h\vec{x}_{02};Y,Y_g\Rb\right\}\nn
  \eea
Assuming that $N^\pi $ falls down at high energy we can neglect the $(N\pi)^2$ - term and one can see that \eq{EQHE} can be re-written in the form:
\bea \label{EQHE1}
&&\frac{\partial N^\pi\Lb x_{10}, b;Y,Y_g\Rb}{\partial Y}\,\,=\\
&=&\,\,\frac{\bas}{2 \pi}\,\int d^2 x_2 \frac{x^2_{10}}{x^2_{02}\,x^2_{12}}\,\left\{ -\Lb \,N^\pi\Lb x_{02}, \vec{b} + \h\vec{x}_{12};Y,Y_g\Rb\,+\,N^\pi\Lb x_{12}, \vec{b} - \h\vec{x}_{02};Y,Y_g\Rb\right. \right.\nn\\
&-& \,\,\left.\left. N^\pi\Lb x_{12}, \vec{b} ;Y,Y_g\Rb\,\Rb \,-\,2\,N^\pi\Lb x_{01};Y,Y_g\Rb\,\right\}\nn
  \eea
  
  This equation can be solved
    using the double Mellin transform for $\tilde{N}$, namely,
    \beq \label{MT}
    N^\pi\Lb \xi, b;Y,Y_g\Rb\,\,=\,\,\int^{\epsilon + i \infty}_{  \epsilon - i \infty} \frac{d \om}{2 \pi i}  \int^{\epsilon + i \infty}_{  \epsilon - i \infty} \frac{d f}{2 \pi i}   \,  e^{ \om Y \,+\,f \xi}\, n^\pi\Lb f, b;\om,Y_g\Rb
    \eeq
    where $\xi\,=\,\ln\Lb x^2_{01}\,\Rb$. It should be stressed that  we need to consider $x^2_{10}\,Q^2_s\Lb Y,b\Rb \,\gg\,1$ since only in this region we can replace $N$ by unity ($Q_s$ is the saturation momentum for $N^\pi$).    
    
   Indeed, for $n^\pi$  we obtain the following equation
    \beq \label{EQHE4}
      \om \,n^\pi\Lb f, b;\om,Y_g\Rb\,\,=\,\,-\bas \Big\{\chi(f) \,n^\pi\Lb f, b;\om,Y_g\Rb\,\,+\,4 \frac{\partial n^\pi\Lb f, b;\om,Y_g\Rb}{\partial f}\Big\}
      \eeq
      where $\chi\Lb f \Rb \,=\,2 \psi(1) - \psi(f) - \psi(1 -f)$ is the BFKL kernel \cite{BFKL} in $f$-representation ($\psi$ is di- gamma function , see formulae {\bf 8.360 - 8.369} in Ref.\cite{RY}).
      
       The solution to \eq{EQHE4} is
       \beq \label{NTIL}
       \tilde{n}^\pi\Lb f, b;\om,Y_g\Rb\,\,=\,\,C(Y_g,b)\,\,e^{ + \frac{\om \,f}{4 \bas} \, + \,\frac{1}{4} \int^f_0 d f' \chi\Lb f'\Rb}
       \eeq       
    Taking Mellin integral of \eq{MT} by the  steepest decent method we obtain the following equations for the saddle points:
    \beq \label{SP}
   \frac{ f_{SP}}{4 \bas}\,\,+\,Y\,=\,0;\,\,\,\,\,
   \,\,\,\frac{\om_{SP} }{4 \bas}\,\,+ \frac{1}{4} \chi\Lb f_{SP}\Rb \,\,+\,\,\xi\, = \,0;
   \eeq
   Taking into account that $\chi\Lb f \Rb\,\,\xrightarrow{f \,\gg\,1}\,\,- 2 \ln f $ we obtain the  solution:

    \beq \label{HESOL}
        N^\pi\Lb \xi, b;Y,Y_g\Rb  \,\,=\,\,\,C(Y_g,b)\,\,e^{ - 4 \bas Y \, \Lb \xi \, -\,\ln \xi\Rb} 
      \eeq
      
      This solution demonstrates that the influence of the interaction in initial and final states are so essential that they lead to decrease of the cross section of the inclusive production of the leading neutrons. This makes the extraction of pion deep inelastic structure from this experiment problematic if not impossible. However, for  more practical  conclusions it is necessary  to solve \eq{EQ} in more realistic kinematic region. For this goal we need to know  the initial condition that we are going to discuss in the next section.
      
\section{Initial conditions}

The problem of the initial conditions actually includes two different issues. The first is to find the expression for the Born Approximation (see \fig{genpic}) for the leading neutron production in DIS for the initial energy. The second one relates to the explicit form of shadowing correction  at the same low energy.
Restricting ourselves by calculation the survival probability we do not need to know the details of Born Approximation which can be found in Ref.\cite{BA}, since they cancels in  the expression for the survival probability, namely,
\beq \label{S2}
S^2\,\,=\,\,\frac{d \sigma_{incl}\Lb \mbox{exact with shadowing corrections}\Rb}{d Y_n \,d^2 p_{n,T}}\Big{/}\frac{d \sigma_{incl}\Lb \mbox{Born Approximation of \fig{genpic}}\Rb}{d Y_n \,d^2 p_{n,T}}     
 \eeq
 where $Y_n $ and $p_{n,T}$ are rapidity and transverse momentum of produced neutron.   
 
 The Born approximation at low energies looks as it is shown in \fig{baga}.
  The expression for this diagram  takes the following form (see Ref. \cite{BA} ):
 \beq \label{BAPH}
 z \frac{d \sigma^{BA}_{p \to n}}{d z d^2 q_T}\,\,=\,\,\frac{1}{16 \pi^2}\,q^2_T\frac{ G^2_{p\pi^+n}\Lb q^2_T\Rb}{\Lb m^2_\pi + q^2_T\Rb^2} \Lb 1 - z\Rb^{ 1 - 2 \alpha_\pi\Lb q^2_T\Rb}\,\frac{4 \pi^2\,\alpha_{em}}{Q^2}\,F^\pi\Lb x_\pi; Q^2\Rb
 \eeq

 where $\alpha_{em}$ is the fine structure constant.  In the rest frame  of proton the longitudinal momentum, carried by pion , is equal to $q_L\,=\,(1 - z)m_N/\sqrt{z}$ and 

  \bea \label{BANOT}
&& x_\pi \,=\,\frac{x}{1 - z}\,=\,\frac{Q^2}{Q^2\,+ \,M^2_X};\,
\,Y_g\,\,=\,\,-\ln(1 - z);\,\,\,\,\alpha_\pi = 
 \alpha' \Lb  m^2_\pi - q^2_T\Rb \,\,\to\,\,\mbox{pion Regge trajectory}
 \eea
\FIGURE[h]{
\begin{tabular}{c }
\epsfig{file=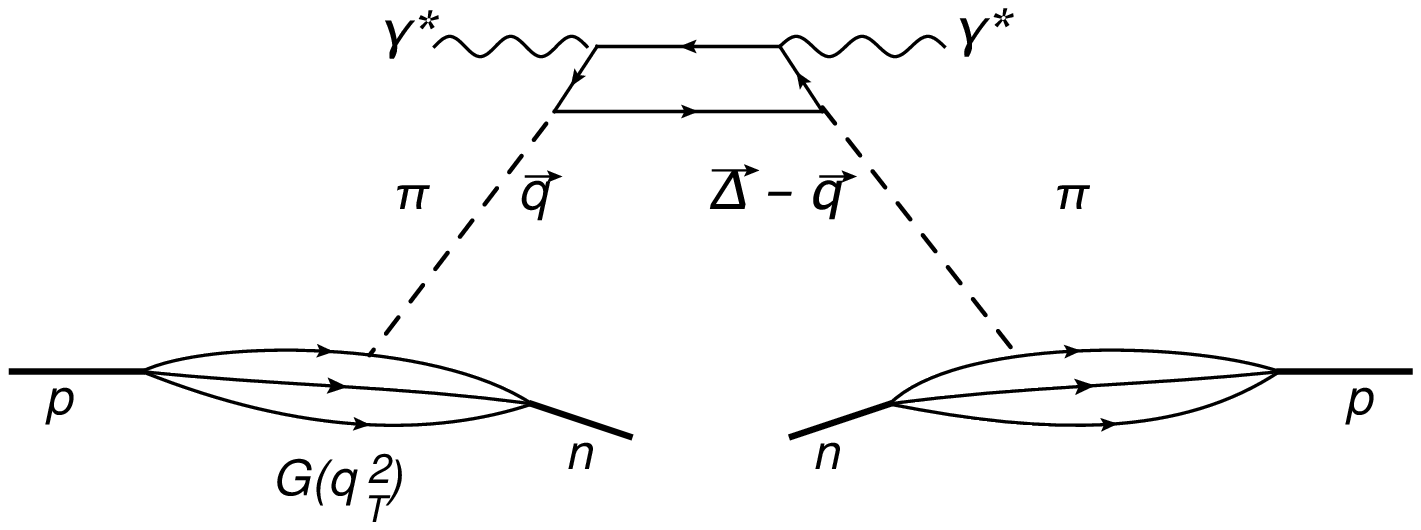,width=70mm}\\
\end{tabular}
\caption{Born Approximation for $\ga^* p$ scattering art low energies. The dashed line describes pion, the solid arrowed line corresponds to quark(antiquark)}
\label{baga}
}
 
  For initial condition we need to take $F^\pi\Lb x_\pi \to 1, Q^2\Rb$ at $x_\pi \to 1$. For calculation of survival probability (see \eq{S2})   we do not need to know the value of $F^\pi$ and even its $x_\pi$ dependence. The only ingredient that we need, has not appeared in \eq{BAPH}. The total cross section corresponds the $\ga^* \pi$ amplitude that is taken at momentum transferred from virtual photon to virtual photon ($\Delta$ in \fig{baga})  to be  equal to zero. For calculation of the survival probability we need to know the $\Delta$ dependence of this amplitude. We explain this fact a little bit below. The expression for such an amplitude takes the following form
  \bea \label{BAPH1}
&&N^\pi\Lb \Delta; Q;Y= Y_g,Y_g\Rb\,\,=\\
&&\,\,\frac{1}{16 \pi^2}\vec{q}_T \cdot (\vec{\Delta} - \vec{q})_T \frac{G_{p\pi^+n}\Lb q^2_T\Rb}{m^2_\pi + q^2_T}\frac{ G_{p\pi^+n}\Lb( \vec{\Delta} - \vec{q})^2_T\Rb }{m^2_\pi +( \vec{\Delta} - \vec{q})^2_T}\Lb 1 - z\Rb^{ 1 - \alpha_\pi\Lb q^2_T\Rb\,-\,\alpha'_\pi \Lb  (\vec{\Delta} - \vec{q})^2_T\Rb}\,\frac{1}{Q^2}\,F^\pi\Lb x_\pi; Q^2;\Delta\Rb\nn
 \eea
 We do not know the dependence of $F^\pi$ on $\Delta$. In the additive quark model (see \fig{baga})  it is natural to assume that $ F^\pi\Lb x_\pi; Q^2;\Delta\Rb\,\,=\,\,   F^\pi\Lb x_\pi; Q^2\Rb  \times G_\pi\Lb \Delta^2\Rb$ where $G_\pi$ is the electro-magnetic form factor of pion. On the other hand 
 the main contribution for scattering at low energy gives the $\rho$-resonance which contribution has no $\Delta$-dependence.

  Integrating  \eq{BAPH1} over $q_T$ and going to impact parameter representation 
 \beq \label{IPR}
 N^\pi\Lb b; Y=Y_g,Y_g \Rb \,\,=\,\,\int \frac{d^2 \Delta_T}{(2 \pi)^2}\,e^{i \vec{\Delta}_T \cdot \vec{b} }N^\pi\Lb \Delta\Rb
 \eeq
 we obtain that
 \beq \label{AB}
N^\pi\Lb b \Rb\,\,=\,\,N_g\Lb Y_g\Rb\,\int d^2 b' \,G_\pi \Lb \vec{b} - \vec{b}'\Rb\,T^2\Lb b' \Rb
\eeq 
 
 ~
    
    where  factor $N\Lb Y_g\Rb $ includes the dependence of $Y_g$($z$) and does not contribute to the calculation of the survival probability (see \eq{S2}). In \eq{AB} $G_\pi\Lb b \Rb$ is the impact parameter image of $G_\pi \Lb \Delta \Rb$ while
    \beq \label{AB1}
    T\Lb b \Rb\,\,=\,\,\frac{\partial }
    {\partial b}\int \frac{d^2 q_T}{(2 \pi)^2}\,e^{i \vec{q}_T \cdot \vec{b}  } \Lb 1 - z\Rb^{\alpha'_\pi q^2_T}   \frac{G_{p \pi^+ n}\Lb q^2_T\Rb}{m^2_\pi + q^2_T}\,\,=\,\,\frac{1}{2 \pi} \int q^2_T d q_T J_1\Lb q_T\, b\Rb\,\frac{G_{p \pi^+ n}\Lb q^2_T\Rb}{m^2_\pi + q^2_T}\Lb 1 - z\Rb^{\alpha'_\pi q^2_T}  
         \eeq
      
   We use $G_\pi\Lb \Delta\Rb$ in the standard form
   \beq \label{GPI}
      G_\pi\Lb \Delta\Rb   \,\,=\,\,\frac{1}{1 + \Delta^2/\mu_\pi^2}   
      \eeq
      where $\mu^2_\pi = 0.6 GeV^2$  is taken from the measured value of the electro-magnetic radius of pion:$ R_\pi = 0.66 \pm 0.01 fm $\cite{RPI}. For $G_{p \pi^+ n}$ form factor there exists a variety 
of different experimental (phenomenological ) information (see Ref. \cite{BA} for details).      We take this form factor to be proportional to electro-magnetic form factor of proton in the spirit of the additive quark model that we pictured in \fig{baga}. It takes a form
\beq \label{GPPIN}
G_{p \pi^+ n}\Lb q_T\Rb\,\,=\,\,g_{p \pi^+ n}\,\,G_p\Lb q_T\Rb\,\,=\,\,\frac{g_{p \pi^+ n}}{\Lb 1 + q^2_t/\mu^2_p\Rb^2}
\eeq
with $\mu^2_p \,=\,0.72 GeV^2$ which corresponds to $R_p = 0.862 \pm 0.012 fm$. In Ref.\cite{BA}
 is suggested to replace 
 \beq \label{REPL}
 G_{p \pi^+ n}\Lb q_T\Rb e^{\alpha'_\pi q^2_T Y)_g} \,\longrightarrow\,\,\frac{1}{1 + q^2_T/\mu^2_{eff}}
 \,\,\,\mbox{with}\,\,\,\,\frac{1}{\mu^2_{eff}}\,\,=\,\,\frac{2}{\mu^2_p}\,\,+\,\,\alpha'_\pi Y_g
 \eeq
  This expression describes correctly the small $q_T$ behaviour which is the most essential feature of the pion exchange.
  
  Using \eq{REPL} we obtain
  \beq \label{SIMLT}
  T\Lb b \Rb\,\,=\,\,\frac{\mu^2_{eff}}{\mu^2_{eff} - m^2_\pi}\,\Big( m_\pi K_1\Lb m_\pi  b     \Rb\,-\,\mu_{eff}\,K_1\Lb \mu_{eff} b\Rb\Big)
  \eeq
 We will use \eq{REPL} in our estimates. 
      
      Now we need to specify the initial condition for BK equation. As it will be seen below we cannot calculate the survival probability without introducing the impact parameter dependence of the scattering amplitude. It is known, that BK equation has a problem with taking into account the large  $b$-dependence of the scattering amplitude \cite{KOWI} and should be
    modified  at large values of $b$.   We suggest a different approach in the spirit of the additive quark model, assuming that we have two different scales inside the proton: the size of the proton and the size of the constituent quark which is much smaller that the size of proton.
    Since the saturation scaler is larger that $1/R_{\mbox{constituent quark}}$ we can integrate all amplitude in the BK-equations  over impact parameter, In this picture the entire $b$-dependence will be concentrated in initial conditions and will be determined by the form factor of proton.  
 In this model we view a proton in the same way as  tritium  in the Glauber approach (see Ref.\cite{LETA}).   
      Such an approach could be relevant only if we know that the radius of interaction between dipole and constituent quark does not increase with energy. In high energy phenomenology based on soft Pomeron approach, the increase of the radius of interaction $R$ with energy looks as follows
      \beq \label{CQM}
      R^2\,= R^2_Q + \alpha'_{P} \ln(s)\,\,<\,\,R^2_p
      \eeq
      where $R_p $ is the size of the proton and $R_Q$ is the size of the constituent quark,   $\alpha'_P$ is the slope of the Pomeron trajectory. Fortunately, some models have recently been discussed with very small value of $\alpha'_{\pom}\, \approx\, 0.01 GeV^{-2}$ \cite{SOFT} and \eq{CQM} could be valid at high energies.
      
 We will fix the initial condition using results from Ref.\cite{AAMQS} in which a perfect fit to HERA DIS data was made in framework of BK equation.     We use the same initial conditions as in this paper, namely,\footnote{Golec-Biernat and Wusthoff (GBW) model being very simple, describe DIS data\cite{GBW} while McLerran-Venugopalan  formula is the correct initial condition in Colour Gluon Condensate approach\cite{MV}}
 \bea 
&&\mbox{GBW model:}\,\,\,\,\, N\Lb x_{10},Y=Y_g\Rb = \sigma_0\,
\Big( 1 \,-\, \exp\Lb - \Lb x^2_{01}\, Q^2_{0 s}\Rb^\ga/4\Rb\Big)\,; \label{IC01} \\
&&\mbox{MV formula:}\,\,\,\,\, N\Lb x_{10},Y=Y_g\Rb = \sigma_0\,
\Lb 1 \,-\, \exp\Lb - \Lb x^2_{01}\, Q^2_{0 s}\Rb^\ga\,\ln\Lb \frac{1}{ x_{10}\,\Lambda_{QCD}}\,+\,e\Rb/4\Rb\Rb\,;\label{IC02} 
\eea

 but we introduce the $b$ dependence considering     that initial saturation momentum $Q^2_{0s}$ depends on
 $b$ in the following way: 
 \beq \label{IC1}
 Q^2_{0s} \,\,\longrightarrow \,\, Q^2_{0s}\Lb b\Rb\,\,=\,\,Q^2_{0s}\,G_p\Lb b\Rb \,\,=\,\,Q^2_{0s}\, m b\,K_1\Lb m b \Rb
 \eeq 
 where $G_p\Lb b\Rb$ is the $b$ image of \eq{GPPIN}. Therefore,  \eq{IC01} and \eq{IC02}  transform into the following expressions
 \bea \label{IC2}
  N\Lb x_{10},Y=Y_g\Rb = \int \,d^2 b \,N\Lb x_{10},Y=Y_g; b \Rb\,\,=\,\,\left\{\begin{array}{l}
  \int d^2 b \Big( 1 \,-\, \exp\Lb - \Lb x^2_{01}\, Q^2_{0 s}\Lb b \Rb\Rb^\ga/4\Rb\Big) \\ \\ \\
\int d^2 b \Big( 1 \,-\, \exp\Lb - \Lb x^2_{01}\, Q^2_{0 s}\Lb b \Rb\Rb^\ga\ln\Lb \frac{1}{ x_{10}\,\Lambda_{QCD}}\,+\,e\Rb/4\Rb\Big) \end{array}
\right.
\eea
  
  Considering \eq{IC01} and \eq{IC2}  at small $x_{01}$ ($x^2_{01}\, Q^2_{0 s} \ll\,1$)
 we see that 
 \beq \label{IC3}
 \sigma_0 \Lb Q^2_{0 s}\Rb^\ga\,\,=\,\,\int d^2  b  \Lb  Q^2_{0 s}\Lb b\Rb\Rb^\ga 
 \eeq
 This equation gives us the value of
 $Q^2_{0s}$ as function of $\sigma_0$ and $m$.  
 In our solution we took two values for $Q^2_{0s}$  in \eq{IC1}: first one,  considering $m$ being  the same as in the electromagnetic form factor of proton $ m^2 = 0.72\,GeV^2$; and the second one, taking $Q^2_{0s}$ in \eq{IC01}  and \eq{IC02}  being equal to $Q^2_{0s}$ in \eq{IC01}(\eq{IC02}) but changing $m$ to satisfy \eq{IC3}.
  Since it turns out that $\ga$ is close to
 1,   we take as a first try \eq{IC1} with $\ga=1$.

  In this case we have simple initial condition for BK equation  in momentum representation for \eq{IC01}
  \beq \label{IC4}
  N\Lb k,Y=Y_g, b\Rb\,\,=\,\,\int x_{01} d x_{01}\, J_0\Lb b x_{01}\Rb \,\frac{ N\Lb x_{01},Y=Y_g,b\Rb }{x^2_{01}}\,\,=\,\,\h \Gamma_0\Lb k^2/Q^2_{0s}\Lb b \Rb\Rb
  \eeq
  where $\Gamma_0$ is incomplete Euler gamma function (see formulae   {\bf 8.25} in Ref.\cite{RY}).  In the case of the initial condition of \eq{IC02} we did not find a simple analytical form and perform the integral of \eq{IC4} numerically.
  
  The easiest way to get the initial condition for $N^\pi$ is to write them in coordinate representation.  Assuming that at large $Q^2$  $ F^\pi_2 = (2/3)F^p_2$ due to the quark counting rules \cite{LF} we can obtain the initial condition for $N^\pi$ in the form
  of \eq{IC01} and \eq{IC02} in which we replace $Q^2_{0s}(b) $ by $(2/3) Q^2_{0s}(b) $. In other words the initial conditions for $N^\pi$   have the form
 \bea \label{IC21}
 && N^\pi_{BA}\Lb x_{10},Y=Y_g; b \Rb\,\,=\\
 &&\,\,N_g(Y_g)\,G_p\Lb b \Rb \int d^2 b' \,G_\pi \Lb \vec{b} - \vec{b}'\Rb\,T^2\Lb b' \Rb\left\{\begin{array}{l}
\Big( 1 \,-\, \exp\Lb - \frac{2}{3}\Lb x^2_{01}\, Q^2_{0 s}\Lb b \Rb\Rb^\ga/4\Rb\Big) \\ \\ \\
 \Big( 1 \,-\, \exp\Lb - \frac{2}{3} \Lb x^2_{01}\, Q^2_{0 s}\Lb b \Rb\Rb^\ga\ln\Lb \frac{1}{ x_{10}\,\Lambda_{QCD}}\,+\,e\Rb/4\Rb\Big) \end{array}\nn
\right.
\eea
  However, we need to multiply these initial values of $N^\pi_{BA} $ by the survival probability since the interaction in the initial state can fill the rapidity gap  for the leading neutron production.  
  
  In the case of \eq{IC1} the survival  probability is equal to
 \bea \label{SP}
 \mbox{For GBW  model:} &~&  S^2 \,\,=\,\,\exp\Lb - \h x^2_{01}\,Q_{0 s}^2\,G_p\Lb b \Rb\Rb;\nn\\
  \mbox{For MV formula:} &~&  S^2 \,\,=\,\,\exp\Lb - \h x^2_{01}\,Q_{0 s}^2\,G_p\Lb b \Rb\ln\Lb \frac{1}{ x_{10}\,\Lambda_{QCD}}\,+\,e\Rb\Rb; 
  \eea
 since due to unitarity constraint such $S^2$ corresponds to probability not to have any inelastic interaction which can spoil the rapidity gap\cite{BJ,GLM1}.
 Therefore the initial condition for $ N^\pi$ in the  coordinate representation has a form
 \bea 
 &&N^\pi\Lb x_{01}, b, Y=Y_g, Y_g\Rb\,\,\,=\,\,\, S^2   N^\pi_{BA} \Lb x_{01}; b;Y=Y_g,Y_g\Rb\,\,\nn\\
 && \mbox{For GBW  model:}\,\,=\,\,N_g \Lb Y_g\Rb\,\Big( 1 \,-\, \exp\Lb - \frac{2}{3}\Lb x^2_{01}\, Q^2_{0 s}\Lb b \Rb\Rb^\ga/4\Rb\Big)\nn\\
 &&~~~~~~~~~~~~~~~~~~~~~\times\,\,  \exp\Lb - \h x^2_{01}\,Q_{0 s}^2\,G_p\Lb b \Rb\Rb \int d^2 b' \,G_\pi \Lb \vec{b} - \vec{b}'\Rb\,T^2\Lb b' \Rb\label{ICCOR1}\\
 &&   \mbox{For MV formula:}= \,N_g \Lb Y_g\Rb\,\Lb 1 \,-\, \exp\Lb - \frac{2}{3} \Lb x^2_{01}\, Q^2_{0 s}\Lb b \Rb\Rb^\ga\ln\Lb \frac{1}{ x_{10}\,
 \Lambda_{QCD}}\,+\,e\Rb/4\Rb\Rb\nn\\
 &&~~~~~~~~~~~~~~~~~~~~~\times \exp\Lb - \h x^2_{01}\,Q_{0 s}^2\,G_p\Lb b \Rb\ln\Lb \frac{1}{ x_{10}\,\Lambda_{QCD}}\,+\,e\Rb\Rb
    \int d^2 b' \,G_\pi \Lb \vec{b} - \vec{b}'\Rb\,T^2\Lb b' \Rb  \label{ICCOR2}
      \eea

  Going to momentum representation we have the  initial condition in the form:
  \beq \label{INMR}
  N^\pi\Lb k, b, Y=Y_g, Y_g\Rb\,\,\,= \,\,\,2 \pi \int x_{01} d x_{01} \,J_0\Lb x_{01}\,k\Rb\, \frac{N^\pi\Lb x_{01}, b, Y=Y_g, Y_g\Rb}
  {x^2_{01}}\,\,\nn\\
  \eeq

  Factor$N\Lb Y_g\Rb  $ is not important for the calculation of the survival probability if we neglect $\Lb N^\pi\Rb^2$-term in      
   \eq{EQ} (see below). However, we need it for the estimates of the accuracy with which we can neglect this term. It is equal to \cite{BA}
  \beq \label{NG}
  N_g\Lb Y_g\Rb \,\,=\,\,\frac{g^2_{p \pi^+ n}}{16 \pi^2}\,e^{- Y_g}\,
  \,\approx \,\,2.2 \,e^{-\,Y_g}
  \eeq
  
   \section{Numerical solution}
  In this section we will discuss numerical solutions of two equation: BK equation and \eq{EQ}, in momentum representation
  \bea\label{MRP}
   N\Lb x_{01}, b, Y\Rb\,\,& = &\,\,x^2_{01}\,\int k dk  J_0\Lb k x_{01}\Rb \,N\Lb k, b, Y\Rb;\nn\\\,\, 
   N^\pi\Lb x_{01}, b, Y, Y_g\Rb &= &x^2_{01}\,\int k dk
  J_0\Lb k x_{01}\Rb\, N^\pi\Lb k, b, Y, Y_g\Rb;
    \eea
  
  In this representation BK equation looks as follows
  \bea \label{BKMRP}
  \frac{\partial N\Lb k, b, Y\Rb }{ \partial Y}\,&=&\,\\
   & & \bas\left\{\int \frac{d^2 k'_\perp }{\Lb \vec{k} - \vec{k}^{\,\prime} \Rb^2_\perp}\,\Big(N\Lb k^{\,\prime}, b, Y\Rb  
  \,-\,\frac{k^2_\perp}{ k'^2_\perp \,+\,\Lb \vec{k} - \vec{k}^{\,\prime} \Rb^2_\perp} \, N\Lb k, b, Y\Rb \Big)\,-\, N^2\Lb k, b, Y\Rb\right\} ;\nn
  \eea 
  
  This equation we solve with the initial condition given by \eq{IC2}.
  
  After finding the solution to \eq{BKMRP} we solve the following equation:
   \bea \label{EQMPR}
\hspace{-0.2cm}&&  \frac{\partial N^\pi\Lb k, b, Y,Y_g\Rb }{ \partial Y}\,=\,\\
   && \bas\left\{\int \frac{d^2 k'_\perp }{\Lb \vec{k} - \vec{k}^{\,\prime} \Rb^2_\perp}\,\Big(N^\pi\Lb k^{\,\prime}, b, Y,Y_g\Rb  
  \,-\,\frac{k^2_\perp}{ k'^2_\perp \,+\,\Lb \vec{k} - \vec{k}^{\,\prime} \Rb^2_\perp} \, N^\pi\Lb k, b, Y,Y_g\Rb\Big) - 4\,N\Lb k, b, Y\Rb\,N^\pi\Lb k, b, Y,Y_g\Rb\right\} ;\nn
  \eea 
  with the initial condition given by \eq{INMR}
  
  \eq{EQMPR} differs from \eq{EQ} in the momentum representation by the $\Lb N^\pi\Rb^2$-term which we neglected in this equation. Indeed, $N^\pi$ in Born approximation is much smaller than $N$ both experimentally and theoretically, since it describes a specific configuration  that contribute to the inclusive cross section given by $N$. This configuration is suppressed by factor $e^{-Y_g}$. The second argument stems from the solution at high energies (see section 2.2) which shows that  $\Lb N^\pi\Rb^2$ turns out to be much smaller than   $ 4\,N\, N^\pi $.
     We proceed with solution of \eq{EQMPR} but  after finding the solution to this equation we will check that the $\Lb N^\pi\Rb^2$-term 
     is small.

     Using the solution to \eq{EQMPR} we can find $S^2$ for dipole  scattering which is equal to
 \bea \label{S2F}
   & \mbox{$S^2$}\,\,\,=\,\,\,\int d^2 b\, N^\pi\Lb k,b,Y,Y_g\Rb\Big{/}\int d^2 b \,N^\pi_{BA}\Lb k,b,Y,Y_g
   \Rb&
     \eea
     where $N^\pi$ is the solution to \eq{EQMPR} while $N^\pi_{BA}$ is given by \eq{BAPH}.

     One can see that for self-consistent  calculations we need to estimate the contribution of the Born Approximation using \eq{AB} as the initial condition,
 solving the linear BFKL equation. 
     \bea \label{EQLIN}
\hspace{-0.2cm}&&  \frac{\partial N^\pi\Lb k, b, Y,Y_g\Rb }{ \partial Y}\,=\,\\
   && \bas\left\{\int \frac{d^2 k'_\perp }{\Lb \vec{k} - \vec{k}^{\,\prime} \Rb^2_\perp}\,\Big(N^\pi\Lb k^{\,\prime}, b, Y,Y_g\Rb  
  \,-\,\frac{k^2_\perp}{ k'^2_\perp \,+\,\Lb \vec{k} - \vec{k}^{\,\prime} \Rb^2_\perp} \, N^\pi\Lb k, b, Y,Y_g\Rb\Big)   \right\}\nn
  \eea
  
  Therefore, we need to solve numerically three equations: \eq{BKMRP}, \eq{EQMPR} and \eq{EQLIN}.
  We notice that using a new variable for GBW initial condition (see \eq{IC01})
    \beq \label{KAPPA}
  \kappa\Lb b \Rb\,\,=\,\,k{\Big/} Q_{0s}\Lb b \Rb
  \eeq
  the initial conditions for these three equations can be written in the following form 
  \bea 
  \mbox{\eq{BKMRP}}\,\,&\longrightarrow&\,\,  N\Lb k,Y=Y_g, b\Rb\,\,=\,\,\h \Gamma_0\Lb \kappa^2\Lb b \Rb\Rb \,;\label{FIC1}\\
   \mbox{\eq{EQMPR}}\,\,&\longrightarrow&\,\,  N\Lb k,Y=Y_g, b\Rb\,\,= 
  N_g\Lb Y_g\Rb\, \int d^2 b' \,G_\pi \Lb \vec{b} - \vec{b}'\Rb\,T^2\Lb b' \Rb \,\nn\\
 & & ~~~~~~~~~~~~~~~~~~~~~~~\times  \int \frac{d r}{r}\Big(1 - e^{-\frac{1}{6}\,r^2}\Big)\,e^{ - \h r^2}\,J_0( \kappa r);\label{FIC2}\\
  \mbox{\eq{EQLIN}}\,\,&\longrightarrow&\,\, N^\pi\Lb k,Y=Y_g, b\Rb\,\,= 
  \,\,\frac{N_g\Lb Y_g\Rb}{Q^2_{0s}\Lb Y_g, b\Rb}\, \int d^2 b' \,G_\pi \Lb \vec{b} - \vec{b}'\Rb\,T^2\Lb b' \Rb \,\,\h \Gamma_0\Lb\frac{2}{3} \kappa^2\Lb b \Rb\Rb ;\label{FIC3}
  \eea   
  
  Since \eq{EQMPR} and \eq{EQLIN} are linear equations with respect to $N^\pi$ we can find the solution to them with the simplified initial conditions
   \bea 
   \mbox{\eq{EQMPR}}\,\,&\longrightarrow&\,\,  N^\pi\Lb k,Y=Y_g, b\Rb\,\,= \,\, \int \frac{d r}{r}\Big(1 - e^{-\frac{1}{6}\,r^2}\Big)\,e^{ - \h r^2}\,J_0( \kappa r)  \,\,\;\label{FIC21}\\
  \mbox{\eq{EQLIN}}\,\,&\longrightarrow&\,\, N^\pi\Lb k,Y=Y_g, b\Rb\,\,= 
\,\h\,\Gamma_0\Lb\frac{2}{3} \kappa^2\Lb b \Rb\Rb;\label{FIC22}
  \eea   
  and the  solutions with the initial conditions of \eq{FIC2} and \eq{FIC3} can be obtained as 
  \beq \label{SOL1}
   N^\pi\Lb k,Y;  b\Rb\,\,=\,\, \,\,\frac{N_g\Lb Y_g\Rb}{Q^2_{0s}\Lb Y_g, b\Rb}\, \int d^2 b' \,G_\pi \Lb \vec{b} - \vec{b}'\Rb\,T^2\Lb b' \Rb \,\,N^\pi\Lb \kappa\Lb b \Rb, Y\Rb\,
 \eeq
 where $N\Lb \kappa\Lb b \Rb, Y\Rb$ is the solution to \eq{EQMPR}  or to \eq{EQLIN} with the initial conditions given by \eq{FIC21} or \eq{FIC22}.

     All equations that we are discussing here are conformal invariant and can be re-written in variables $\vec{\kappa}$ and $\vec{\kappa}'$ in stead of $\vec{k}$ and $\vec{k}' $. Hence actually we need to solve the system of three equations (\eq{BKMRP}, \eq{EQMPR} and \eq{EQLIN}) with the initial conditions of \eq{FIC1}, \eq{FIC21} and \eq{FIC22} to find $N\Lb \kappa, Y\Rb$  and $N^\pi\Lb \kappa,Y\Rb$.   In spite of simplicity of \eq{FIC1}.\eq{FIC21} and \eq{FIC22} they have clear shortcoming since cannot reproduce a correct behaviour  accordingly to operator product expansion at large values of $k^2$. Indeed, they fall down exponentially instead of  $1/k^2$. Therefore, we can trust  these equations only at $k^2 \leq Q^2_s$. Trying to find a compromise between simplicity and rigorousness we chose the initial condition for $N^\pi$ in the form
  \beq \label{ICNUM1}
  N^\pi_{BA} \Lb x_{01}, b; Y=Y_g,Y_g\Rb\,\,=\,\, \frac{2}{3} \,x^2_{01}\,\ln\Lb\frac{1}{x_{01} \Lambda_{QCD}}\Rb
  \eeq
  which is the expansion of \eq{IC21} for McLerran-Venugopalan formula at  small values of $x_{01}$.
  We can only treat $x_{01} < 1/Q_s$ with this initial condition but since $Q_s$ for low energy in DIS with a pion is small we believe that we can  use \eq{ICNUM1} as the first approximation. In our equation for $N^\pi$ the solution of BK equation is essential only for $x_{01} \leq Q_s$ and, therefore, we can use the GBW model for the initial condition. The same situation in the initial condition for $N^\pi_{BA}$.
  Finally, we use
  \bea
  \mbox{BFKL equation:}&  ~~~~~&   \frac{2}{3}x^2_{01}\,Q^2_{0s}\ln\Lb \frac{1}{x_{01}\,\Lambda_{QCD}}\Rb ; \label{NUM22}\\
  \mbox{our equation:}&  ~~~~~&\frac{2}{3}x^2_{01}\,Q^2_{0s}\ln\Lb \frac{1}{x_{01}\,\Lambda_{QCD}}\Rb   \exp\Big( - x^2_{01}\,Q^2_{0s}/4\Big) ; \label{NUM23}   
 \eea
   These equations take the following form in the momentum representation:
   \bea
  \mbox{BK equation:}&  ~~~~~&  \h \Gamma_0\Lb k^2/Q^2_{0s}\Rb ; \label{NUM31}\\
  \mbox{BFKL equation:}&  ~~~~~&   \frac{1}{3}\,Q^2_{0s}/k^2; \label{NUM32}\\
   \mbox{our equation:}&  ~~~~~&\frac{1}{3}\,1/Q^2_{0s}\Lb -\Gamma_0\Lb - \frac{k^2}{Q^2_{0s}}\Rb
  \,  \exp\Big( - k^2/Q^2_{0s}\Big)\Rb;\label{NUM33}
 \eea 

 The expression in \eq{NUM33} follows from the following simple calculations:
 \bea\label{CAL}
 \ln\Lb1/ x^2_{01}\Rb & \xrightarrow{\mbox{Fourrier transform}} & \frac{1}{k^2};\nn\\
 (x^2)^n \ln \Lb1/ x^2_{01}\Rb & \xrightarrow{\mbox{Fourrier transform}} & \Lb- \frac{1}{k^2} \frac{d^2}{(d \ln k^2)^2}\Rb^n \frac{1}{k^2} \,\,=\,\,(-4)^n (n!)^2 \frac{1}{(k^2)^{n+1} }
 \eea
 and from formula   {\bf 8.357} of Ref.\cite{RY}.
 
 We need to find the pion deep inelastic structure function, using $N^\pi$ and $N^\pi_{BA}$,
 for calculating the value of the survival probability. In the  dipole approach the cross section for the virtual photon with the pion can be expressed through the amplitude of the dipole-pion interaction in the following way for massless quarks:
 \beq \label{DIXS}
 \sigma_{\gamma^* \pi}\Lb Q, Y\Rb \,\,=\,\,\int d^2 x_{01}\,|\Psi\Lb Q;x_{01}\Rb |^2\,\sigma_{\mbox{dipole-$\pi$}}\Lb x_{01},Y\Rb\,\,=\,\,\int d^2 x_{01}\,|\Psi\Lb Q;x_{01}\Rb|^2\,\int d^2 b N^\pi\Lb x_{01},Y; b\Rb\
 \eeq
 where \cite{WAFU}
 \beq \label{WF}
 |\Psi\Lb Q;x_{01}\Rb|^2\,\,=\,\,\frac{2 N_c \alpha_{em}}{\pi}\sum_f Z^2_f \,\int d z [ z^2 + (1-z)^2]\,\bar{Q}^2 K_1^2\Lb \bar{Q} x_{01}\Rb 
 \eeq
 where $Z_f$ is the fraction of  the electrical charge $\alpha_{em}$ carried by quark of flavour $f$;  $Q$ is the photon virtuality and $\bar{Q}^2 = Q^2 z (1 - z)$ ; $z$ is the fraction of photon energy carried by the quark an $N_c$ is the number of colours. Using \eq{MRP} \eq{DIXS} can be re-written in the form
 \beq \label{DIMR}
  \sigma_{\gamma^* \pi}\Lb Q, Y\Rb \,\,= \,\, \frac{2 N_c \alpha_{em}}{\pi\,Q^2}\sum_f Z^2_f\,\,\int  k\, d k\, \Phi^2_{\gamma^*}\Lb Q,k\Rb \int d^2 b  \,N^\pi\Lb k ,Y; b \Rb
  \eeq
  where
  \bea \label{PHI}
 \Phi^2_{\gamma^*}\Lb \tau=\frac{Q}{k}\Rb \,&=&\,Q^4\,\int d^2 x_{01}\int d z\,z (1-z)\, [ z^2 + (1-z)^2] \,x^3_{01} \,J_0\Lb k x_{01}\Rb\ K_1^2\Lb \bar{Q}\, x_{01}\Rb \\
 &= &8\,\tau^4\int d z\, z (1 - z)\, [ z^2 + (1-z)^2]  \frac{(1 - \tilde{\kappa}^2)
  \sqrt{1 + 4 \tilde{\kappa}^2}\,+\,8 \tilde{\kappa}^2 ( 1  +  \tilde{\kappa}^2)\mbox{ ArcCsch}\Lb 2 \tilde{\kappa}\Rb}{ (1 + 4   \tilde{\kappa}^2)^2     \sqrt{1 + 4 \tilde{\kappa}^2}}\nn
  \eea
  where $\tilde{\kappa} = \bar{Q}/k = \tau \sqrt{z(1-z)}$.
  
   Using $N^\pi\Lb \kappa,Y\Rb$ , \eq{DIMR} and \eq{PHI} the final expression for the survival probability (see \eq{S2F}) has the following form

 {\Large
       \bea \label{S2FF}
 & \mbox{$S^2$}\,\,\,=\,\,\,\frac{\int d^2 b \, \int d^2 b' \,G_\pi \Lb \vec{b} - \vec{b}'\Rb\,T^2\Lb b' \Rb\,\int^{\infty}_{\Lambda_{QCD} /Q_{0s}(b)}\kappa d \kappa\, \Phi^2_{\gamma^*}\Lb Q/\Lb Q_{0s}(b)\, \kappa\Rb\Rb\,N^\pi\Lb \mbox{\normalsize\eq{EQMPR}};\, \kappa , Y\Rb}{\int d^2 b \, \int d^2 b' \,G_\pi \Lb \vec{b} - \vec{b}'\Rb\,T^2\Lb b' \Rb\,\int^{\infty}_{\Lambda_{QCD} /Q_{0s}(b)}\kappa d \kappa\, \Phi^2_{\gamma^*}\Lb Q/\Lb Q_{0s}(b)\, \kappa\Rb\Rb\,N^\pi\Lb \mbox{\normalsize\eq{EQLIN}};\,\kappa, Y\Rb}&
     \eea   
}

In \eq{S2FF} we use the infrared cuttoff $k = \Lambda_{QCD}$ since for smaller $k$ we cannot use \eq{ICNUM1} for the dipole amplitude.
  \begin{figure}
  \begin{tabular}{c c}
\epsfig{file=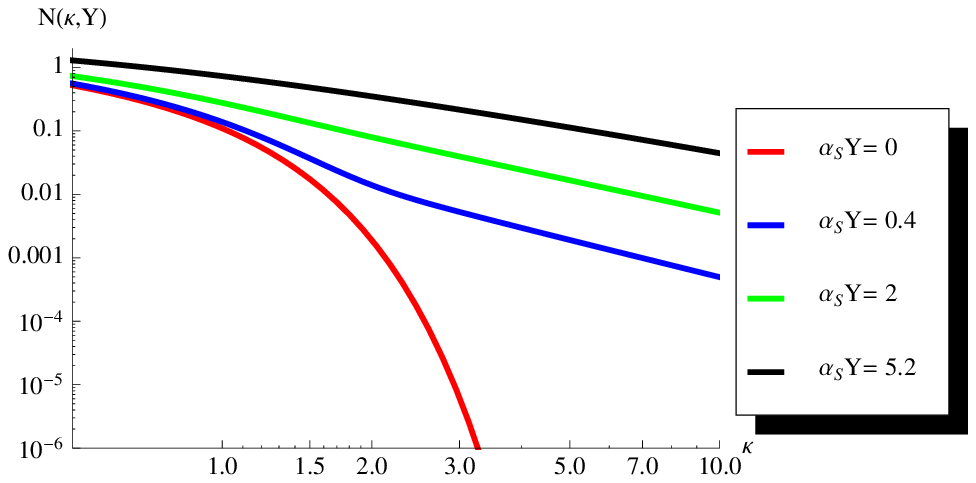,width=80mm} & \epsfig{file=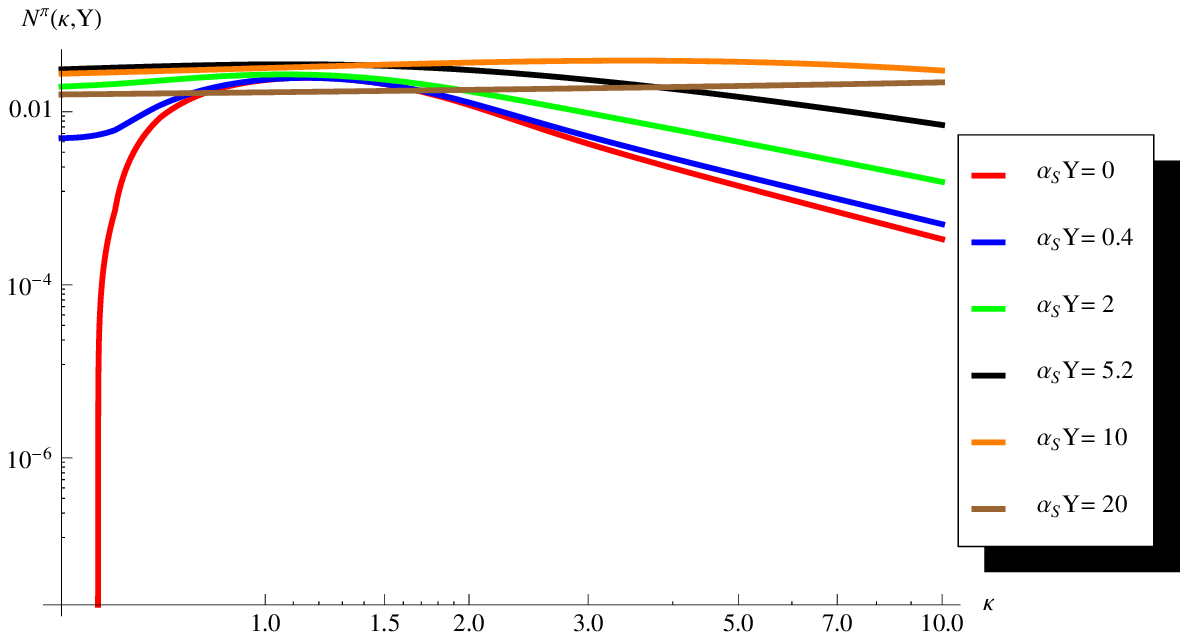,width=80mm} \\ 
\fig{sol}-a & \fig{sol}-b\\
\epsfig{file=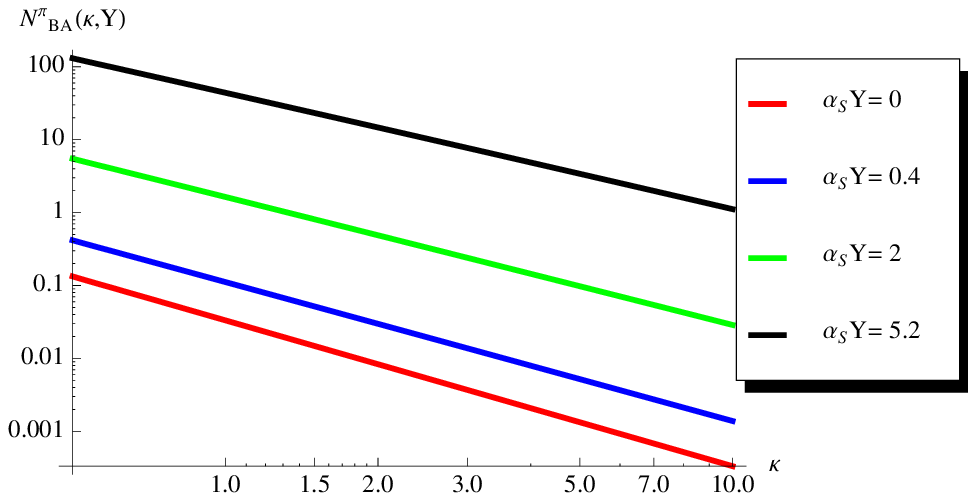,width=80mm} &{\begin{minipage}{80mm}{\vspace{-30mm} Solutions to equations at different rapidities:
\fig{sol}-a  shows the solution to the Balitsky-Kovchegov  equation with the initial solution of \protect \eq{FIC1};
the solution to new equation (see \protect\eq{EQMPR} ) with the initial condition of \protect\eq{FIC2} is plotted in \fig{sol}-b, while the solution of the BFKL equation with the initial condition of \protect\eq{FIC3} is shown in \fig{sol}-c. The value of $Y_g$ is taken to be equal to 3. }
\end{minipage}}
\\
\fig{sol}-c &  \\
\end{tabular}
\caption{Solutions to equations as function of $\kappa$ at different energies}
\label{sol}
\end{figure}     
     
The solutions to \eq{BKMRP},\eq{EQMPR} and \eq{EQLIN} with the initial conditions given by \eq{FIC1},\eq{FIC2} and \eq{FIC3} are shown in \fig{sol}. One can see that the solution of the linear BFKL equation (see 
\fig{sol}-c ) steeply increases with energy while the solution to the Balitsky-Kovchegov equation shows only a mild increase with rapidity ( see
 \fig{sol}-a). The solution to the new equation increases with energy  but starts to  fall down at very high energies  (see \fig{sol}-c).   From \fig{sol} one can see that the term $\Lb N^\pi\Rb^2$ turns out to be much smaller than the term $4 \,N\, N^\pi$  in \eq{EQ} for small $\kappa$ ( $ \kappa \leq 2$). For large $\kappa$ this term is much smaller than the linear term in the equation ( $N^\pi \,\ll\,\Lb N^\pi\Rb^2$). Therefore, we can conclude {\it a posteriori} that we can neglect $\Lb N^\pi \Rb^2$-term in \eq{EQ} and reduce this equation to \eq{EQMPR} which has been solved.

 The main result of the paper: the estimates for the survival probability ( see \eq{S2FF}) , is shown in \fig{sp} for two different choice of the dependence of the saturation momentum on $b$ given by   \eq{IC1}. The first one shown in solid lines, corresponds to \eq{IC1}   where $m$ is chosen the same as in the electromagnetic 
form factor of the proton and value of $Q^2_{0s}$ is found from \eq{IC3}. The second choice was to fix the value of $Q^2_{0s}$ to be the same as in Ref.\cite{AAMQS} but the value of $m$ is determined by \eq{IC3}. The survival probability is shown by dotted lines in \fig{sp}. The qualitative features are seen in \fig{sp}: the value of the survival probability is small and its decreases with the growth of energy (rapidity).  The smallness stems mostly from the steep increase of the solution to the linear BFKL equation while to the considerable decrease contribute two factors: the increase of the solution to linear equation and decrease of the solution to the new equation (see \eq{EQMPR}). The third interesting feature is that the value of the survival probability at high energies (large values of rapidity)  does not depend on the initial conditions for $\kappa \geq 2$. It is worthwhile mentioning that the errors that stem from the different initial condition turns out to be smaller than      $\delta S^2/S^2  \leq 0.1$ for any value of $\kappa$.
\FIGURE[h]{
\centerline{\epsfig{file=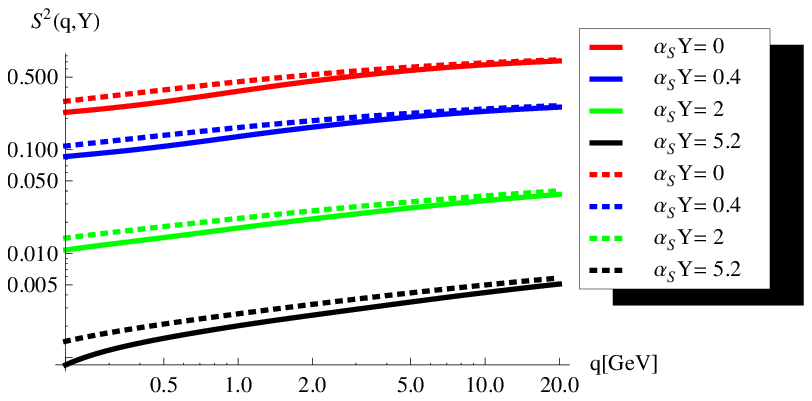,width=110mm}}

\vspace{-1.5cm}
\caption{The values of the survival probabilities  for leading neutron production in DIS as a function of  the photon virtuality ($q$)  at different energies (rapidities).The solid and dotted line corresponds to different choice of the value for the typical mass in \eq{IC1}. The solid lines describe the initial condition with $m$ chosen to be the same as in the electromagnetic for factor of proton and the value of $Q^2_{0s}$ is determined by \eq{IC3}, while dotted lines present the different choice: $Q^2_{0s}$ is taken to be  the same as in Ref.\protect\cite{AAMQS} while $m$ is the solution of \eq{IC3}.}
\label{sp}
}
   \section{Conclusions}
   The main result of this paper is \eq{EQ} (see \fig{eq}) for the cross section of the inclusive production of leading neutrons in DIS. This equation stems from the direct generalization of the approach developed in Ref.\cite{KL} for the diffractive production in DIS. The asymptotic solution to this equation as well as  the numerical solution to \eq{EQ} shows that the survival probability defined in \eq{S2}, is small and steeply falls down with energy. We believe that these features of the survival probability are general and does not depend on a particular process that we consider. Being the first theoretical attempt to calculate the survival probability this paper shows that the survival probability could be as small as $10^{-3}$ at high energies.
   
   However, the numbers we need to take with considerable cautions, since these estimates depend crucially on the assumed impact parameter dependence of both the DIS structure function and  the Born approximation for the leading neutron inclusive cross section. Unfortunately, the phenomenological analysis of DIS data based on Balitsky-Kovchegov equation   (see Ref. \cite{AAMQS}) was performed neglecting the impact parameter dependence. Therefore, to obtain the reliable estimates we need to re-visit the DIS data and re-do the analysis using Baitsky-Kovchegov equation the impact parameter depending initial conditions.
   
  Therefore, we consider this paper as only the first step to the reliable estimates for the experimentally measured cross section.   The small value of the survival probability as well as its energy dependence make difficult  the task of extraction of the deep inelastic structure function for pion, measuring the spectrum of the leading neutron.

                 \section*{Acknowledgements}
We thank Boris Kopeliovich for the instructive discussion on Born Approximation for leading neutron production in DIS and for providing us a possibility to read Ref.\cite{BA} before publication.
This work was supported in part by the  Fondecyt (Chile) grant 1100648.


\end{document}